\documentclass[%
reprint,
superscriptaddress,
onecolumn,
 amsmath,amssymb,
 aps,
]{revtex4-2}

\usepackage[latin9]{inputenc}
\usepackage{amssymb}
\usepackage{graphicx}
\usepackage{amsfonts}

\begin{document}
\title{Accuracy of the Gross-Pitaevskii Equation in a Double-Well Potential}
\author{Asaad R. Sakhel}
\affiliation{Department of Physics, Faculty of Science, Al-Balqa Applied
University, Salt 19117, Jordan}
\author{Robert J. Ragan}
\affiliation{Department of Physics, University of Wisconsin-La Crosse, WI
54601 USA}
\author{William J. Mullin}
\affiliation{Department of Physics, University of Massachusetts, Amherst,
Massachusetts 01003 USA}
\date{\today}

\begin{abstract}
The Gross-Pitaevskii equation (GPE) in a double well potential produces
solutions that break the symmetry of the underlying non-interacting
Hamiltonian, i.e., asymmetric solutions. The GPE is derived from the
more general second-quantized Fock Schr$\ddot{\mathrm{o}}$dinger
equation (FSE). We investigate whether such solutions appear in the
more general case or are artifacts of the GPE. We use two-mode analyses
for a variational treatment of the GPE and to treat the Fock equation.
An exact diagonalization of the FSE in dual-condensates yields degenerate
ground states that are very accurately fitted by phase-state representations
of the degenerate asymmetric states found in the GPE. The superposition
of degenerate asymmetrical states forms a cat state. An alternative
form of cat state results from a change of the two-mode basis set.
\hfill \break\break
Bose condensate, cold atoms, quantum fluids, nonlinear dynamics, Gross-Pitaevskii equation.
\end{abstract}
\maketitle

\section{Introduction}

The solution of the Gross-Pitaevskii equation (GPE) is a single-particle
function that describes Bose condensation \cite{String}. The equation
has had considerable success, perhaps especially in treating atoms
in traps. A particularly interesting trap is the double well \cite{Milb}-\cite{Mal}
which allows tunneling; one can find multiple phenomena, including
bifurcations to asymmetric states \cite{Masiello1}-\cite{Mal},
self-trapping \cite{Milb,Masiello2,Ostro,Coullet}, etc. Symmetry breaking has been observed experimentally in several types
of systems \cite{Green,KevExp,Hamb,ZiboldExp}. The GPE can
be derived from the second-quantized Fock Hamiltonian by replacing
the field operator with a classical function. In doing this, one neglects
the non-commutation of the field operator as a small effect of order
of the inverse number of particles. A possible question about various
effects shown by the GPE is whether those same or similar effects
also occur in the original Fock Schr$\ddot{\mathrm{o}}$dinger equation
(FSE) or are they simply artifacts of the GPE. In this paper we center
this question on the nonlinear bifurcations that result in asymmetric
quantum condensate states. We ask how those states might also appear
in the more general Fock description. 

There have been extensions of the GPE from the Fock Hamiltonian applicable
to trapped gases, for example, the Hartree-Bogoliubov approximation
(HBA) by Fetter \cite{Fetter}, Ruprecht et al \cite{Rup}, Griffin \cite{Griff},
and Popov \cite{Popov}; also see the summaries in the book by Pethick and Smith \cite{PethSmit}.
These treatments replace the field operator with a c-number condensate
function $\Psi(x)$ plus a correction operator $\phi(x)$ that represents
excitations. This quantity is kept to second order in the Fock
Hamiltonian and the resulting quadratic Hamiltonian is diagonalized
in terms of Bose excitation operators by solving Bogoliubov equations.
Our analysis involves a two-mode approximation to the field operator;
we keep all orders of the creation and annihilation operators yielding
fourth order tunneling terms (neglected in the HBA) that allow dual-condensate states representing asymmetric many-particle states in the
Fock eigenstates.

The second quantized Fock Hamiltonian in one dimension (in unitless
form) is 
\begin{equation}
H=\int dx\Psi^{\dagger}(x)\left[-\frac{\partial^{2}}{\partial x^{2}}+V_{ex}(x)\right]\Psi(x)+\frac{1}{2}\int dxdx^{\prime}\Psi^{\dagger}(x)\Psi^{\dagger}(x{}^{\prime})W(x-x^{\prime})\Psi(x^{\prime})\Psi(x),\label{eq:Fock}
\end{equation}
with external potential $V_{ex}(x)$, interatomic potential $W(x-x^{\prime})=\eta\delta(x-x^{\prime})$,
and $\int dx\Psi^{\dagger}(x)\Psi(x)=\hat{N}$, the number operator.
Positions are measured in terms of a length $a,$ and energies in
\begin{equation}
\varepsilon=\frac{\hbar^{2}}{2ma^{2}},
\end{equation}
and time in $\hbar/\varepsilon.$ Replace the field operator $\Psi(x)$
by a classical variable $\Phi(x)$ to find the energy
\begin{equation}
E=\int dx\Phi^{*}(x)\left[-\frac{\partial^{2}}{\partial x^{2}}+V_{ex}(x)+\frac{\eta}{2}\left|\Phi(x)\right|^{2}\right]\Phi(x),
\end{equation}
where $\int dx\left|\Phi(x)\right|^{2}=N$. The time dependent GPE
can be derived by varying the energy with respect to $\Phi^{*}$ to
get
\begin{equation}
i\frac{\partial\Phi}{\partial t}=\left(-\frac{\partial^{2}}{\partial x^{2}}+V_{ex}(x)+\eta\left|\Phi\right|^{2}\right)\Phi.
\end{equation}
Inserting an exponential time dependence, $\Phi(x,t)=e^{-i\mu t}\sqrt{N}\psi(x),$
yields the time-independent GPE in the form
\begin{equation}
\left[-\frac{d^{2}}{dx^{2}}+V_{ex}(x)+\eta N\left|\psi(x)\right|^{2}\right]\psi(x)=\mu\psi(x),\label{eq:GPE}
\end{equation}
with $\int dx\left|\psi(x)\right|^{2}=1$.

In a recent article \cite{RSM} we solved this equation for
the ``box-delta'' potential
\begin{equation}
V_{ex}(x)=\left\{ \begin{array}{cc}
\infty, & \left|x\right|\ge1\\
\gamma\delta(x), & \left|x\right|<1
\end{array}\right.,
\end{equation}
for which we found analytic solutions of the equation including bifurcations
to asymmetric wave functions. Here we want to test the accuracy of
the GPE relative to the FSE by treating the box-delta potential approximately
using a two-mode model. Two-mode models have been used extensively
in studying Bose-Einstein condensation \cite{Milb}-\cite{Masiello1},\cite{Rag},\cite{Ostro},\cite{Coullet},\cite{AGK}-\cite{Sac2}
and we will find it useful here as well. We begin with a two-mode
variational treatment of the GPE, followed by the same method applied
to the FSE. 

\section{Variational treatment of GPE\label{sec:Variational-treatment-of}}

\begin{figure}[h]
\includegraphics[width=6in]{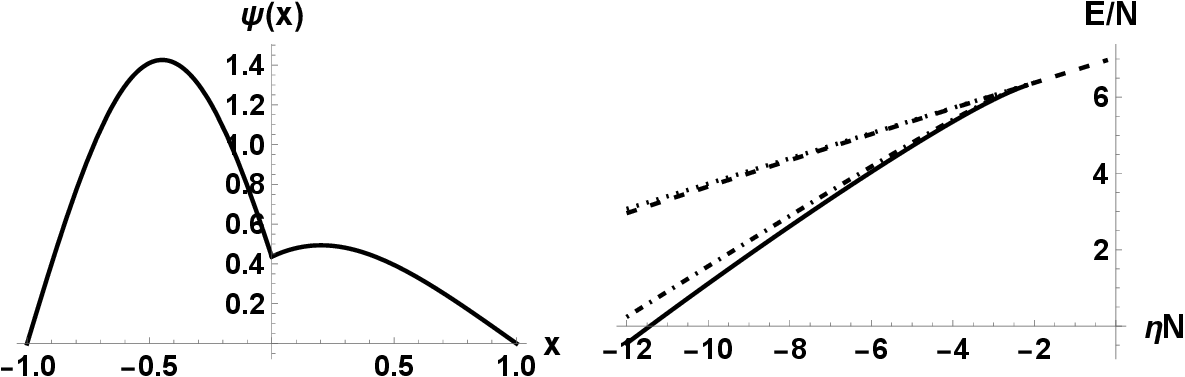}

\caption{{[}Left{]} Asymmetrical variational wave function for $\gamma=10$
and $\eta N=-4.2$ with variational parameter $u=0.8615.$ {[}Right{]}
Bifurcation to asymmetric states: (Dashed line) The exact energy per
particle \cite{RSM} as a function of interaction $\eta N$ for the symmetric
state, and (solid line) the exact asymmetric state. (Dotted line--barely
above the exact dashed line) the variational symmetric state; and
(dash-dotted line) that for the asymmetric state. The variational
bifurcation occurs at $\eta N=-2.3$ compared to the exact result
\cite{RSM} at $-2.07.$}

\label{fig1}
\end{figure}

The two states that make up a variational trial wave function are
the lowest two states of the non-interacting single particle Hamiltonian,
that is, Eq. (\ref{eq:GPE}) with $\eta=0.$ These states are
\begin{eqnarray}
\psi_{0}(x) & = & A\left(\sin k\left|x\right|+\frac{2k}{\gamma}\cos kx\right),\label{eq:psi0}\\
\psi_{1}(x) & = & \sin\pi x.\label{eq:psi1}
\end{eqnarray}
In $\psi_{0}$ the wave number $k$ satisfies 
\begin{equation}
\tan k=-\frac{2k}{\gamma}.
\end{equation}
The energies of these two ideal gas states are, respectively, 
\begin{eqnarray}
e_{0} & = & k^{2},\label{eq:e0}\\
e_{1} & = & \pi^{2}.\label{eq:e1}
\end{eqnarray}
so that $\psi(\pm 1)=0$.
It has been shown in Refs. \cite{Sac1,Sac2} that for any double-well potential the two-state superposition using the lowest states
of the noninteracting Hamiltonian, as in Eqs. (\ref{eq:psi0}) and
(\ref{eq:psi1}), leads to an exponentially small error for the GPE
in the semiclassical limit, which in our case is $\gamma\rightarrow\infty$ \cite{RSM}.
We take $\gamma=10$ throughout, for which the two-mode approximation
is found to be quite accurate \cite{RSM}.  The trial wave function is 
\begin{equation}
\psi(x)=u\psi_{0}(x)+v\psi_{1}(x),
\end{equation}
which gives energy 
\begin{equation}
E=u^{2}e_{0}+v^{2}e_{1}+\frac{\eta N}{2}\left(\chi_{40}u^{4}+\chi_{04}v^{4}+6\chi_{22}u^{2}v^{2}\right),
\end{equation}
where
\begin{equation}
\chi_{nm}=\int dx\psi_{0}^{n}(x)\psi_{1}^{m}(x).
\end{equation}
With interactions, the symmetric state ($u=1)$ is the lowest energy
until $\eta N\le-2.3$ at which point a superposition of symmetric
and antisymmetric states has lower energy, forming an asymmetric state.
The asymmetric state has its main peak on one or the other side of
the double well (see Fig \ref{fig1}). In the exact calculation of
Ref. \cite{RSM} the bifurcation point is at $\eta N=-2.07$.
Fig \ref{fig1} shows the energy as a function of $\eta N$ in both
the exact and variational treatments. 

This trial wave function gives a second (degenerate) asymmetric solution
as well, with $v$ now having the opposite sign from the first case
and main peak on the opposite side.

\section{Two-mode treatment of the Fock Hamiltonian}

To test the accuracy of the GPE in producing asymmetric states, we
treat the full FSE, also by the two-mode method.

\subsection{The energies}

In the Fock Hamiltonian of Eq. (\ref{eq:Fock}), we substitute the
field operator
\begin{equation}
\Psi(x)=a_{0}\psi_{0}(x)+a_{1}\psi_{1}(x),
\end{equation}
with $a_{i}$ the annihilation operator for particles in the state
$i$. In the double-well potential, with, say, an attractive
interaction one might expect that it might be preferable to use localized
states, with particles gathering together in one well or the other.
That is, a more physical basis might be 
\begin{eqnarray}
\psi_{+} & = & \frac{1}{\sqrt{2}}\left(\psi_{0}+\psi_{1}\right),\label{eq:psi+}\\
\psi_{-} & = & \frac{1}{\sqrt{2}}\left(\psi_{0}-\psi_{1}\right).\label{eq:psi-}
\end{eqnarray}

However, this question of ``best'' physical basis is the
whole point of this section. The set of Eqs. (\ref{eq:psi0}) and
(\ref{eq:psi1}) and that of Eqs. (\ref{eq:psi+}) and (\ref{eq:psi-})
are connected by a unitary transformation and so description in terms
of one can be transcribed quickly into description in terms of the
other. Moreover, we look for a physical description in an even better
basis than either. We will however return to this localized basis
later in Sec. \ref{subsec:Cat-states}. It turns out to be somewhat
easier mathematically to work with the symmetric-antisymmetric pair
for which the tunneling term in $H$ is simpler. 

We take matrix elements of the Hamiltonian in the occupation
states $\left|n_{0},n_{1}\right\rangle $. The diagonal part of the
energy is
\begin{equation}
\frac{H_{0}}{N}=\varepsilon_{0}\frac{n_{0}}{N}+\varepsilon_{1}\frac{n_{1}}{N}+\frac{\eta N}{2}\frac{1}{N^{2}}\left[\chi_{40}n_{0}\left(n_{0}-1\right)+\chi_{04}n_{1}\left(n_{1}-1\right)+4\chi_{22}n_{0}n_{1}\right]\label{eq:diag}
\end{equation}
with 
\begin{equation}
n_{0}+n_{1}=N.
\end{equation}
There is also an off-diagonal part of the Hamiltonian, which is
\begin{equation}
\frac{H^{\prime}}{N}=\frac{\eta N}{2}\frac{\chi_{22}}{N^{2}}\left(a_{0}^{\dagger2}a_{1}^{2}+a_{1}^{\dagger2}a_{0}^{2}\right).\label{eq:offdiag}
\end{equation}
This connects states $\left|n_{0},n_{1}\right\rangle $ with $\left|n_{0}+2,n_{1}-2\right\rangle $ and $\left|n_{0}-2,n_{1}+2\right\rangle $.   As we will see it is this
term that allows the existence of many-body dual-condensate states
that are the equivalent of the asymmetric states in the solution of
the GPE.

\begin{figure}[h]
\centering \includegraphics[width=3in]{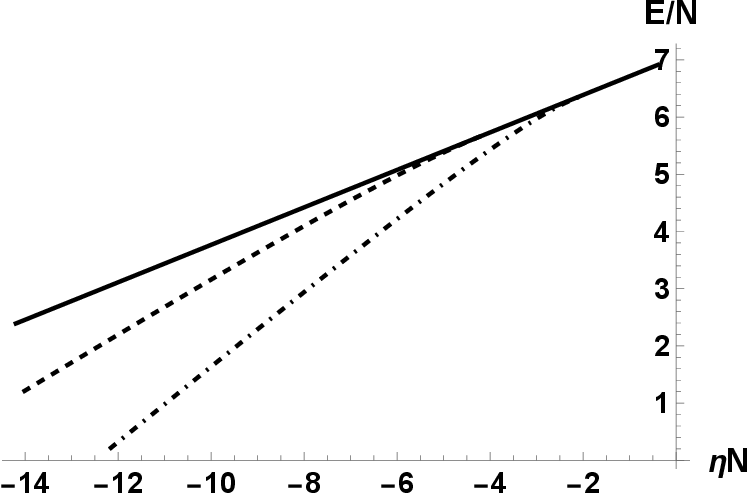}

\caption{Energies per particle for the Fock Hamiltonian with $N=100$: (Solid
line) Energy per particle of a single condensate in the symmetric
state versus $\eta N$ [Eq. (\ref{eq:diag}) with $n_{0}=N$, $n_{1}=0$].
(Dashed line) Energy per particle for a dual condensate in the two
lowest states using only the diagonal energy of Eq. (\ref{eq:diag}).
(Dash-dotted line) Energy per particle for the exact diagonalization
of the Hamiltonian in dual-condensate states including the tunneling
term of Eq. (\ref{eq:offdiag}). The energy per particle calculated using the GPE is almost  identical to this energy. }

\label{Two-states}
\end{figure}

We proceed in two steps,
first treating just the diagonal energy, and then including the off-diagonal
terms as well. In these calculations we take $N=100.$  Minimizing the diagonal terms of Eq. (\ref{eq:diag}) with respect
to $n_{0}$ to introduce the possibility of having two condensates,
we find that for $\eta N>-4.1$ all the particles condense into the
symmetric state; however for $\eta N<-4.1$ a dual condensate has
lower energy. This is shown in Fig. \ref{Two-states} in the symmetric
state energy and the dual-condensate diagonal energy. 

The off-diagonal tunneling term mixes the dual-condensate states to
give a wave function of the form
\begin{equation}
\Gamma =\sum_{n}C_{n} \left|n,N-n\right\rangle .\label{eq:dual}
\end{equation}
Because of the quartic nature of the tunneling, states of odd $n$
will be connected to other odd $n$ states and even are connected
to even. If the interaction satisfies $\eta N<-2.1,$ the energy now
becomes lower than both the symmetric state and the diagonal energy,
and we then have a superposition of dual condensates. We also find
that the lowest states of the full Fock Hamiltonian are doubly degenerate.
The plot of this lowest energy is shown as the dash-dotted curve in
Fig. \ref{Two-states}. 

It is interesting to compare the GPE asymmetric variational energy
of Fig. \ref{fig1} (dash-dotted curve) with the Fock energy of
Fig. \ref{Two-states} (dash-dotted curve); it is almost exactly the
same, which seems a remarkable coincidence, since the Fock Hamiltonian
has so much extra physics, including dual condensates and tunneling. 

\subsection{The states\label{subsec:The-states}}

\begin{figure}[h]
\includegraphics[width=6in]{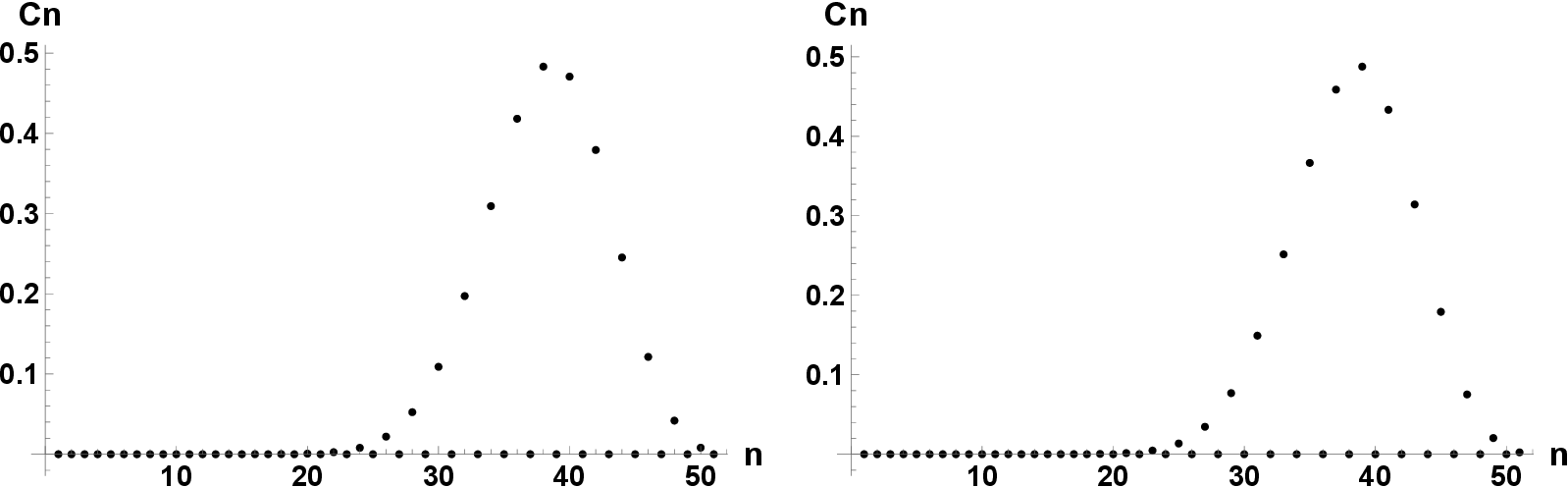}

\caption{Eigenfunctions $\Phi_{a}$ and $\Phi_{b}$, of the Fock Hamiltonian for $N=50$ at $\eta N=-4.2.$
The two states are degenerate ground states with energy per particle
5.34. The first (left, call it $\Phi_{a}$),  has all coefficients $C_{n}$ with $n$ odd
vanishing, while the second (right, $\Phi_{b}$, ) has $C_{n}$ with $n$ even vanishing. }

\label{figStatesA}
\end{figure}

When the interaction is sufficiently negative one gets dual-condensate
states of the form of Eq. (\ref{eq:dual}) and the lowest energy is
degenerate. This is certainly reminiscent of the bifurcation in the
GPE, in which asymmetric degenerate states became those with the lowest
energy. But what are the states here? We pick a particular interaction
parameter $\eta N=-4.2$ with $N=50.$ The energy per particle is
5.34. The two states, $\Phi_{a}$ and $\Phi_{b}$, corresponding to this energy are shown in Fig.
\ref{figStatesA}. At first glance the two seem identical, however,
the first has all coefficients $C_{n}$ with odd $n$ vanishing, while
the second has those with even $n$ vanishing. The two states are
clearly orthogonal. 

It is useful to compute the one-body density matrix (OBDM) for these
states. From the equation
\begin{equation}
\rho_{1}=\left\langle a_{i}^{\dagger}a_{j}\right\rangle 
\end{equation}
with wave functions of the form of Eq. (\ref{eq:dual}) we have the
matrix elements
\begin{eqnarray}
\left\langle a_{0}^{\dagger}a_{0}\right\rangle  & = & \sum_{n}\left|C_{n}\right|^{2}n;\quad\left\langle a_{1}^{\dagger}a_{1}\right\rangle =\sum_{n}\left|C_{n}\right|^{2}\left(N-n\right);\nonumber \\
\left\langle a_{0}^{\dagger}a_{1}\right\rangle  & = & \sum_{n}C_{n+1}^{*}C_{n}\sqrt{\left(n+1\right)\left(N-n\right)};\nonumber \\
\left\langle a_{1}^{\dagger}a_{0}\right\rangle  & = & \sum_{n}C_{n-1}^{*}C_{n}\sqrt{n\left(N-n+1\right)}.\label{eq:OBDM}
\end{eqnarray}
Clearly, if alternating coefficients vanish, $\rho_{1}$ is diagonal and
the state can be fragmented. This is the case for the wave functions
shown in Fig. \ref{figStatesA}. We find for each of these two the result
\begin{equation}
\rho_{1}=\left[\begin{array}{cc}
38.42 & 0\\
0 & 11.58
\end{array}\right].\label{eq:Fragrho}
\end{equation}
Indeed each state is fragmented and the natural orbitals are the symmetric
and antisymmetric basis states. 

\begin{figure}[h]
\includegraphics[width=3in]{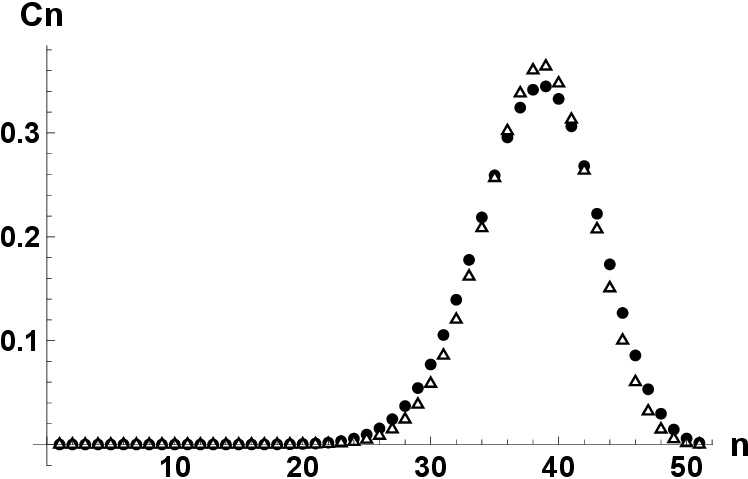}$\quad\quad$\includegraphics[width=3in]{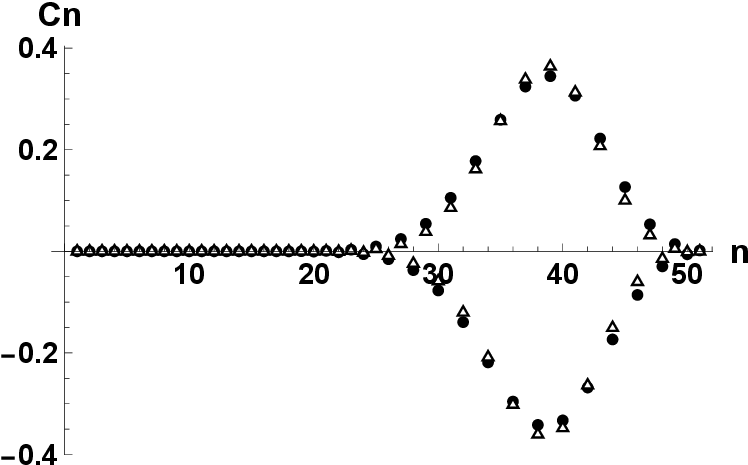}

\caption{Rotated states (Black dots): States (left: $\Psi_{1}$ and right: $\Psi_{2}$) that are linear combinations {[}Eq.
(\ref{eq:rot}){]} of the degenerate ground states of the Fock Hamiltonian,
chosen to make the largest single condensates. (Open triangles): Fitted coefficients Eqs.(\ref{eq:D+}) and (\ref{eq:D-}) of pure condensate asymmetric states. }

\label{figStatesB}
\end{figure}

However these states are not the only basis set possible and it is
useful to see whether we can find ``pure states,'' that is, states
that have nearly 100\% condensate occupancy. We used \textit{Mathematica} to
diagonalize the Hamiltonian and that has some mechanism built in that
provides the particular basis used with degenerate states. Call the
states produced there $\Phi_{a}$ and $\Phi_{b}$, as shown in Fig. \ref{figStatesA}. Then we can produce
a new basis set for this ground state by a rotation through angle
$\theta$:
\begin{eqnarray}
\Psi_{1} & = & \Phi_{a}\cos\theta+\Phi_{b}\sin\theta,\nonumber \\
\Psi_{2} & = & \Phi_{a}\sin\theta-\Phi_{b}\cos\theta.\label{eq:rot}
\end{eqnarray}
We choose $\theta$ to maximize the condensate number in a single
state. We find by trial new states with $\theta=0.78540$ that have
the largest OBDM eigenvalues. These states are shown in Fig. \ref{figStatesB}.
The OBDMs are now given by
\begin{equation}
\rho_{1}=\left[\begin{array}{cc}
38.42 & \pm21.08\\
\pm21.08 & 11.58
\end{array}\right]\label{eq:rho1}
\end{equation}
with eigenvalues
\begin{eqnarray}
N_{1} & = & 49.987,\nonumber \\
N_{2} & = & 0.013.\label{N0}
\end{eqnarray}

If the numerical states shown are related to the basic asymmetric
condensate states found with the GPE, then we should be able to fit
them with binomial coefficients generated from the pure asymmetric
state and its mirror image:
\begin{eqnarray}
\xi_{+} & = & \frac{1}{\sqrt{N!}}(ua_{0}^{\dagger}+va_{1}^{\dagger})^{N}\left|0\right\rangle ,\label{eq:base}\\
\xi_{-} & = & \frac{1}{\sqrt{N!}}(ua_{0}^{\dagger} - va_{1}^{\dagger})^{N}\left|0\right\rangle ,\label{eq:mirror}
\end{eqnarray}
with $\upsilon=\sqrt{1-u^{2}}.$ (These are asymmetrical because the single-particle state $(ua_{0}^{\dagger}\pm va_{1}^{\dagger})$ is a sum of a symmetric and an antisymmetric wave function.)  These two states are nearly orthogonal; we will discuss this aspect below. When expanded the coefficients in each are, respectively,
\begin{eqnarray}
D_{n}^{+} & = & \sqrt{\frac{N!}{n!(N-n)!}}u^{n}v^{N-n},\label{eq:D+}\\
D_{n}^{-} & = & \sqrt{\frac{N!}{n!(N-n)!}}u^{n}(-v)^{N-n}.\label{eq:D-}
\end{eqnarray}
We compare these forms with the numerical plots in Fig. \ref{figStatesB}
in the open triangles where a least square fit gives $u=0.87653$.
The fit is good but not exact. The energies $E/N$ of the states are
as follows: \textit{Mathematica} states [Fig. \ref{figStatesA}]: 5.3391; the fitting states $\xi_{\pm}$:
5.3402, which is a bit higher, as it should be. It is interesting
that the condensate [Eq. (\ref{N0})] is not pure; there seems to be a small occupation of an excited state. This state, with the same
$u$ and $v$, is
\begin{equation}
\xi_{3}=\frac{1}{\sqrt{N!}}(va_{0}^{\dagger}-ua_{1}^{\dagger})^{N}\left|0\right\rangle .
\end{equation}
This state is orthogonal to $\xi_{+}$ and has energy 6.874; it is
an excited state. The coefficients corresponding to this state oscillate like the right plot of Fig. \ref{figStatesB}, but with the peak at small $n=12$ rather than large $n$.

The conclusion here is that the physically most reasonable combination
of states $\psi_{0}$ and $\psi_{1}$ is not the localized set $\{\psi_{+},\psi_{-}\}$
of Eqs. (\ref{eq:psi+}) and (\ref{eq:psi-}) but is the set of asymmetric
states using parameters $u$ and $v$, as used in Eqs. (\ref{eq:base})
and (\ref{eq:mirror}). The states of Eqs. (\ref{eq:base}) and (\ref{eq:mirror}), which represent degenerate condensates in asymmetric states, are eigenfunctions of the Fock Hamiltonian in the two-modes approximation. As such this feature is reminiscent of what occurs in the GPE. 

\subsection{The fitting parameters}

We found values of the rotation angle $\theta$ and the value of $u$
that allowed a fit of the least-fragmented states. Let's see why we
got the numbers we did. States like Eqs. (\ref{eq:base}) and (\ref{eq:mirror})
in the form 
\begin{equation}
\left|u,v,N\right\rangle =\frac{1}{\sqrt{N!}}(ua_{0}^{\dagger}+va_{1}^{\dagger})^{N}\left|0\right\rangle 
\end{equation}
are phase states. (If $u=\left|u\right|e^{i\phi_{u}}$ and $v=\left|v\right|e^{i\phi_{v}}$,
then the ``phase'' is $\phi=\phi_{v}-\phi_{u}$.) Phase states with
different phases become orthogonal when $N\rightarrow\infty$ \cite{MKL}.
In the case of $\xi_{+}$ and the mirror state $\xi_{-}$ we find
that $\left\langle \xi_{+}|\xi_{-}\right\rangle =(u^{2}-v^{2})^{N}=3\times10^{-14}$.
On the other hand, the numerical states that we generated are exactly
orthogonal. A phase state is a kind of eigenfunction of the destruction
operator \cite{MKL}:
\begin{eqnarray}
a_{0}\left|u,v,N\right\rangle  & = & u\sqrt{N}\left|u,v,N-1\right\rangle ,\\
a_{1}\left|u,v,N\right\rangle  & = & v\sqrt{N}\left|u,v,N-1\right\rangle .
\end{eqnarray}
With these equations it is easy to show that the OBDMs of $\xi_{\pm}$
are 
\begin{equation}
\rho_{\pm}=N\left[\begin{array}{cc}
u^{2} & \pm uv\\
\pm uv & v^{2}
\end{array}\right].\label{eq:rhouv}
\end{equation}
Numerically, \textit{Mathematica} found Eq. (\ref{eq:rho1}) so that comparison
with Eq. (\ref{eq:rhouv}) gives $Nu^{2}=38.42$ or that $u=0.8766.$
The least squares fit gave $0.87653$. 

The transformation Eqs. (\ref{eq:rot}) between states is unitary,
so that, if our original $Mathematica$ Fock states are $\Phi_{a}$
and $\Phi_{b}$, we can write
\begin{eqnarray}
\Phi_{a} & \approx & \xi_{+}\cos\theta+\xi_{-}\sin\theta,\nonumber \\
\Phi_{b} & \approx & \xi_{+}\sin\theta-\xi_{-}\cos\theta.
\end{eqnarray}
But from Eqs. (\ref{eq:D+}) and (\ref{eq:D-}) we see that the simple
sum $\xi_{+}+\xi_{-}$ will have odd numbered coefficients vanishing,
while the difference $\xi_{+}-\xi_{-}$ will have even numbered coefficients
zero. That implies $\cos\theta=\sin\theta$ or $\theta=\frac{\pi}{4}=0.78540$
whereas our minimization of the small condensate led to $0.78540$, in agreement.

\subsection{A peculiar condensate}

A condensate of the general form of that of Eq. (\ref{eq:base}) is
rather standard: all $N$ particles are in the same single-particle
state. In this case that state is a superposition of two other states
and it appears to the $N$th power. The state represented
by the OBDM of Eq. (\ref{eq:Fragrho}) is fragmented with 38 particles
in the ground symmetric state $\psi_{0}$ [Eq. (\ref{eq:psi0})]
and 12 in the antisymmetric states $\psi_{1} $ [Eq. (\ref{eq:psi1})]. However, the dual condensation
is not in the state $\left|n_{0},n_{1}\right\rangle =\left|38,12\right\rangle ,$
but is in a superposition of $\left|0,50\right\rangle $, $\left|2,48\right\rangle ,$$\cdots$$\left|50,0\right\rangle $
with varying coefficients. This form of fragmentation is not original
to this paper; for example, it appears in Refs. \cite{AJP}, \cite{Baym},
and \cite{HoCio} but we thought the form unusual enough to be worth
pointing out. 

\subsection{Cat states\label{subsec:Cat-states}}

The states $\Psi_{1}$ and $\Psi_{2}$ of Eqs. (\ref{eq:rot}), given
their approximation by $\xi_{+}$ and $\xi_{-}$ of Eqs. (\ref{eq:base}) and (\ref{eq:mirror}), are made up of states
localized mostly on one side of the double well or the other ($ua_{0}^{\dagger}\pm va_{1}^{\dagger})$.
As such, their sum 
\begin{equation}
\Psi_{+}=\frac{1}{\sqrt{2}}\left(\Psi_{1}+\Psi_{2}\right)\approx \frac{1}{\sqrt{2}}\left(\xi_{+}+\xi_{-}\right)\approx \Phi_{a}
\end{equation}
is a sum of macroscopically distinct states and so qualifies as a
``cat state''. Its two components are orthogonal to one
another and so the cat state has the same energy as each of the degenerate
components. Ref. \cite{Mah} also considered the possibility of constructing
a cat state out of asymmetrically localized states of the GPE. 

\begin{figure}[h]
\includegraphics[width=6in]{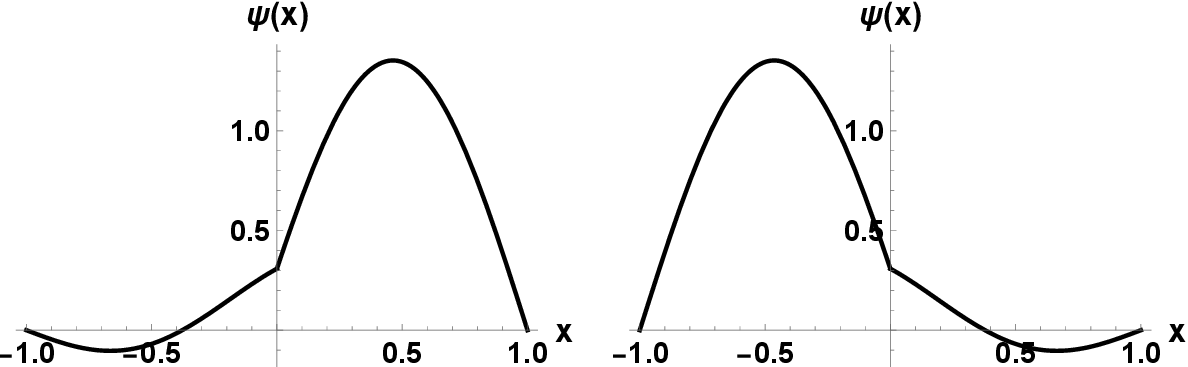}

\caption{The states corresponding to Eqs. (\ref{eq:psi+}) and (\ref{eq:psi-}):
(Left) $\psi_{+}=(\psi_{0}+\psi_{1})/\sqrt{2}$; (Right) $\psi_{-}=(\psi_{0}-\psi_{1})/\sqrt{2}$.}

\label{figBobstates}
\end{figure}

There is another type of cat state that can be considered
here, namely a cat state in dual-condensate space; that is, those
that have double peaks in the distribution of coefficients $C_{n}$
in the states of the form of Eq. (\ref{eq:dual}). These have been
considered in, for example, Refs. \cite{AJP,Baym,HoCio}. When we
expanded the Fock state in a two-mode approximation we used the ground
$\psi_{0}$ and first excited state $\psi_{1}$ of the non-interacting
potential. We found these resulted in combinations that involved asymmetric
states. Suppose we expand directly in asymmetric states by
considering the basis mentioned in Eqs. (\ref{eq:psi+}) and (\ref{eq:psi-}).
These functions are plotted in Fig. \ref{figBobstates}; they are
clearly asymmetric. However, they are also orthogonal unlike most
asymmetric mirror functions. 

\begin{figure}[h]
\includegraphics[width=3in]{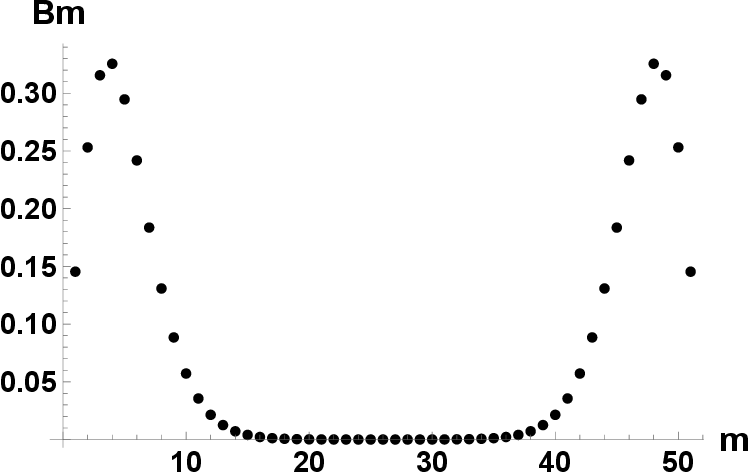}$\quad\quad$\includegraphics[width=3in]{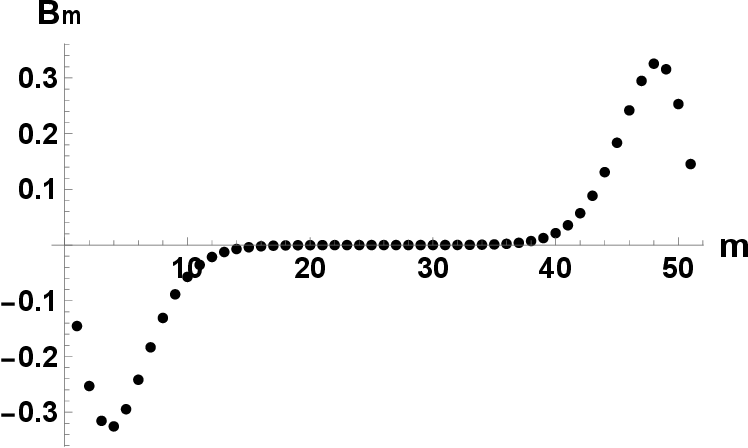}

\caption{Cat states in dual-condensate space gotten by transforming from the
$\left\{ a_{0},a_{1}\right\} $ basis to the $\left\{ a_{+},a_{-}\right\} $
basis of Eqs. (\ref{eq:a+}) and (\ref{eq:a-}). }

\label{figNewCats}
\end{figure}

The operators corresponding to these states are
\begin{eqnarray}
a_{+} & = & \frac{1}{\sqrt{2}}\left(a_{0}+a_{1}\right),\label{eq:a+}\\
a_{-} & = & \frac{1}{\sqrt{2}}\left(a_{0}-a_{1}\right). \label{eq:a-}
\end{eqnarray}
Suppose we make a transformation from dual-condensate states $\left|n,N-n\right\rangle _{01}$
to $\left|m,N-m\right\rangle _{\pm}$, corresponding respectively
to $\left\{ a_{0},a_{1}\right\} $ and $\left\{ a_{+},a_{-}\right\} $
so that the wave function is
\begin{equation}
\Psi=\sum_{n}C_{n}\left|n,N-n\right\rangle _{01}=\sum_{m}B_{m}\left|m,N-m\right\rangle _{\pm},
\end{equation}
where 
\begin{equation}
B_{m}=\sum_{n}A_{mn}C_{n}.
\end{equation}
The transformation matrix is found to be
\begin{equation}
A_{mn}=\frac{1}{2^{N/2}}\sum_{k}\frac{\sqrt{n!m!(N-n)!(N-m)!}(-1)^{N-n-m+k}}{k!(n-k)!(m-k)!(N-n-m+k)!},\label{eq:A}
\end{equation}
where the sum on $k$ (because of the factorials) goes from $\max(0,n+m-N)$
to $\min(n,m).$ When we apply this transformation to the states we
found in Sec. \ref{subsec:The-states}, i.e., those shown in Fig. \ref{figStatesA}, we
find the double-peaked cat states shown in Fig. \ref{figNewCats}. A
particular $m$ value corresponds to $m$ particles in the right well
and $N-m$ particles in the left well. But the true eigenstate is
a linear combination of such states, with peaks symmetricaslly (or
antisymmetrically) at small and large $m$ values, giving cat states.

We can say a bit more about the peaks in the cat states shown here.
Suppose we expand the state of Eq. (\ref{eq:base}) in the set $\left\{ a_{+},a_{-}\right\} $ by using the transformations of Eqs. (\ref{eq:a+}) and (\ref{eq:a-}).
With $u=0.877$ that corresponded approximately to an asymmetric eigenstate
of the Fock Hamiltonian at $\eta N=-4.2$. Plotting the expansion coefficient
of Eq. (\ref{eq:base}) in the set $\left\{ a_{+},a_{-}\right\} $ produces
the right-most peak of the cat state. When we expand Eq.(\ref{eq:mirror})
we get the left-most peak of the cat. Thus each peak represents one
of the many-body asymmetric states found with the Fock Hamiltonian. 

Reference \cite{AJP} showed that if one made measurements one particle
at a time where each measurement determined which of the condensate
states the particle occupied, in this case $\psi_{+}$ or $\psi_{-}$,
then after a rather small number of measurements the cat state would
collapse to just one of the peaks. Is such a cat state, in this case
involving asymmetric states, more capable of being formed than other
kinds? Could asymmetric states aid in making cat states? 

\section{Self-consistent dual condensates}

Cederbaum and coworkers \cite{Ceder1-1} (and others \cite{Masiello1,Masiello2})
have derived self-consistent equations for systems of multicomponent
condensates, called the multi-configurational Hartree theory for bosons
(MCHB). Originally they derived equations from just the diagonal terms
in the Fock Hamiltonian, but later generalized that to use all the
terms, including off-diagonal tunneling terms. The single-condensate
version of MCHB is just the GPE itself. In our two-mode approximation
for the two-condensate Fock Hamiltonian we kept the trial states the
same, the symmetric and antisymmetric lowest states of the noninteracting
potential. In MCHB, the states will change with each change of interaction
parameter and in some cases might become asymmetric states themselves.
A further interesting study would be to consider this type of two-mode
analysis and its connection to the asymmetric states. The MCHB  method 
is capable of treating more than two condensate states as well. We reserve 
such studies for future work. 

\section{Discussion}

We have used a two-mode analysis to compare results derived from the
Fock Hamiltonian to the results using just the GPE. As preparation,
we made a two-mode variational treatment of asymmetric states in the
box-delta potential, which are a reasonably accurate representation
of the bifurcation to the asymmetric states for sufficiently negative
interaction parameter $\eta N.$ The states are pair degenerate with
the wave function peak on one side of the well or the other. 

The GPE yields a single-condensate wave function, having an asymmetric solution
with a degenerate mirror function. In the two-mode treatment of the
Fock $H$, it becomes possible to have dual-condensate states. In
terms of these states, of the form $\left|n\right\rangle \equiv\left|n,N-n\right\rangle $,
there is a diagonal energy as well as an off-diagonal energy representing
tunneling between the $\left|n\right\rangle $ states. The use of
just the diagonal energy leads to a single energetically preferred
dual state $\left|n\right\rangle $ favored over the symmetric single-condensate state below a critical negative interaction strength. However,
inclusion of the tunneling leads to states of the form $\sum_{n}C_{n}\left|n\right\rangle $,
i.e., superpositions of dual-condensates states. Below a critical
interaction parameter value, the states initially produced by the
numerical method used have alternating $C_{n}$ values vanishing either
with odd $n$ or with even $n.$ These two states are degenerate and
have one-body density matrices that are diagonal with fragmented condensates
in the original symmetric and antisymmetric states. These condensates
involve superpositions of odd (or even) numbers of particles in each
of the two basis states. In another representation the asymmetric
states are the natural orbitals with almost all particles falling
into one of those states or the other. There is a small depletion
there with a 0.03\% occupation of another state. Our results demonstrate
that the GPE yields a quite accurate representation of the more general
Fock Schr$\ddot{\mathrm{o}}$dinger equation, with asymmetry occurring
in a very similar form in each. 

Of the numerical states approximated by $\xi_{+}$ and $\xi_{-}$,
one is physically localized more on one side of the well; the opposite
on the other side. The sum of these, one of the states in Fig. \ref{figStatesA},
is a cat state, a sum of two macroscopically distinct states; but
one of precisely the same energy as that of each of the two components,
since the components are orthogonal. A more interesting cat state
was found in dual-condensate space by changing the basis set of the
two-mode analysis. However, it is not clear how asymmetric states
could be used to form cat states experimentally.

\end{document}